\documentclass[preprint]{elsarticle} %TODO: submit with horrible "review" option

\usepackage{lineno,hyperref}
\modulolinenumbers[5]

\journal{Journal of Systems and Software}
\usepackage[acronym, nomain, nonumberlist, style=super]{glossaries}
\usepackage{graphicx}
\usepackage{color}
\usepackage{listings}
\usepackage{caption}
\usepackage{enumitem}
\usepackage{tabularx}
\usepackage{tikz}
\usepackage{tikz-qtree}
\usepackage{tikz}
\usetikzlibrary{arrows,trees,shapes,automata,positioning}
\usepackage{dot2texi}
\usepackage{todonotes}
 \presetkeys{todonotes}{inline}{}
\usepackage{mathtools}

\DeclarePairedDelimiter\abs{\lvert}{\rvert}
\usepackage[export]{adjustbox}
\usepackage{rotating}

\usepackage{url}
\usepackage{booktabs}
\usepackage[ddmmyyyy]{datetime}

\usepackage[autostyle=true, style=american]{csquotes}

\newcommand{\avz}{Avi\v{z}ienis }
\newcommand{\weblink}[1]{\newline \tiny{\url{#1}}}

\newcommand{\interpretation}[1]{\textsc{Interpretation: }{#1}}

\newtheorem{researchquestion}{Research Question}

\newacronym{sms}{SMS}{Systematic Mapping Study}

\begin{document}

\nocite{*}

\title{The landscape of software failure cause models}
\author{Lena Feinbube}
\author{Peter Tr\"oger}
\author{Andreas Polze}

\begin{abstract}
%% TODO: rewrite

The software engineering field has a long history of classifying software failure causes. Understanding them is paramount for fault injection, focusing testing efforts or reliability prediction. Since software fails in manifold complex ways, a broad range of software failure cause models is meanwhile published in dependability literature.
We present the results of a meta-study that classifies publications containing a software failure cause model in topic clusters. Our results structure the research field and can help to identify gaps. We applied the systematic mapping methodology for performing a repeatable analysis.
%Our categorization relies on a refined version of the Laprie/\avz dependability terminology.

We identified 156 papers presenting a model of software failure causes. Their examination confirms the assumption that a large number of the publications discusses source code defects only. Models of fault-activating state conditions and error states are rare. Research seems to be driven mainly by the need for better testing methods and code-based quality improvement. Other motivations such as online error detection are less frequently given. Mostly, the IEEE definitions or orthogonal defect classification is used as base terminology. The majority of use cases comes from web, safety- and security-critical applications.
\end{abstract}

\begin{keyword}
systematic map, mapping study, software, fault, bug, defect, model
\end{keyword}

\maketitle

\section{Introduction}\label{introduction}

More and more aspects of everyday life are supported, enriched, controlled and managed by software systems. This is not only true for commodity devices, but increasingly also for safety-critical environments. Examples exist in the automotive domain, in medical environments and in home automation.
%TODO: citation
This trend seems to be irreversible, which makes software dependability a problem of increasing relevance. Understanding and formalizing the underlying causes and mechanisms leading to software failure is therefore an ever more important task. Such knowledge can be applied in automated testing, reliability modelling, developer education, bug statistics or fault tolerance mechanisms.

A practical example for the usefulness of failure cause models is testing by \emph{fault injection}. Fault injectors artificially insert faults and error states into a running system. This raises the question of which faults to inject, and when. The related work calls this problem \emph{fault representativeness} \cite{faultrepresentativeness}. It demands a well-defined and sufficiently realistic model of erroneous software states and their originating causes.

However, an unfortunate characteristic of the field of software dependability is the gap between theoretical research and industrial practice.
During co-operation with different industry partners, we have learned that practicioners, working with complex dependability-critical software systems, are often constrained by nondisclosure agreements and therefore reluctant to share their experiences and models of software failures.
Due to the realization that every software flaw poses a threat to security, real world software failure data is usually not made public.
On the other hand, a vast body of academic research has been published, which attempts to understand and classify the ways in which software fails.

While this gap between theory and practice is hard to close entirely, the problem can be mitigated by establishing a common terminology base and improving the accessibility of research results.
With this meta-study of software failure cause models, we strive to structure existing theoretical research and thus make it better accessible to practicioners.

In our description of the current landscape of published failure cause models, we collect relevant publications in a repeatable way using the \gls{sms} methodology. We first define a tag categorization scheme which can be used to characterize all relevant publications in the field. Subsequently, the result of our study is a clustering of the research field according to eight different categories.

Starting from the fact that failure causes are denoted in different ways in the literature (e.g. \enquote{defect model}, \enquote{error model}, \enquote{bug model}), we first define a common terminology foundation in Section \ref{terminology}. Subsequently, we present our primary research questions in Section \ref{rqs} and the research methodology we applied in Section \ref{sms}. The tags and tag categories used for classification are presented in Section \ref{tagnames}. Finally, the results are analyzed and discussed in Section \ref{results}.

\section{Terminology}\label{terminology}

Models for describing the failure behaviour of software have been widely discussed in different software engineering areas. We experienced that published models are not only diverse in their purposes and application domains, but also in the terminology they use. They freely mix up problematic issues of development, requirements specification, and production phase. Static software artefacts, such as source code and configuration files, are often treated in the same way as runtime phenomena leading to fault activation, such as environment and race conditions.
For fruitful discussions of software failure causes, it is therefore essential to establish a base terminology first. In this section, we briefly describe the terminology basis we use -- presented in more detail in \cite{terminology}. 

Our terminology of \emph{software failure causes} -- a term we chose deliberately to be unambiguous -- relies on the terminology model by Laprie/\avz. It is driven by three basic concepts: failure, error, and fault. We use \enquote{software failure cause} as an umbrella term for these concepts. With the words of Laprie/\avz \cite{avizienis2004basic},

\begin{quotation}
A software \textbf{failure} is an event that occurs when the delivered service deviates from correct service from user perspective. An \textbf{error} is that detectable part of the system state that may cause a subsequent failure. A \textbf{fault} is the adjudged or hypothesized cause of an error.
\end{quotation}

While the Laprie/\avz terminology is dominant in classical engineering domains, it seems to have mixed adoption in software-centric dependability research. Publications in the software engineering domain often have their own abstraction of error causes and error states. For example, \enquote{error} in the IEEE software engineering glossary specification \cite{radatz1990ieee} maps to the \enquote{fault} concept in the Laprie/\avz model. 

In the Laprie/\avz model, the term \enquote{fault} denotes error-causing internal structural deficiencies and error-causing external influences at the same time. We adjust this interpretation in order to tailor it to the world of software failure causes: We interpret faults only as design imperfections here, similarly to \cite{grottke08} and \cite{laski}, by using a slightly modified version of the fault definition by Pretschner et al. \cite{pretschner2013generic}:

\vspace{1em}
\emph{A software fault is a minimal set of code deviations from correct code, such that the execution of the deviating code can trigger an error.}
\vspace{1em}

We expect the deviation from correct code to be minimal, meaning that all parts of it are mandatory for having a potential error cause. The triggering of a transition to an error state is not mandatory, but possible when the according state conditions are given. %Figure \ref{fig:terminology_model} shows our terminology as an automaton visualization.

The circumstances of fault activation are widely acknowledged to be crucial, but somehow form an elusive fuzzy aspect in software. As an extension to the Laprie/\avz terminology, we therefore introduce the notion of \emph{fault activation conditions} to better grasp the software-specific patterns which lead from a disabled fault to an error state. They depend on the internal system state only.

Based on the above mentioned terminology, we distinguish between four types of models:

\begin{itemize}
\item The \textbf{fault model}, describing code-based static defects in the program;
\item The \textbf{(fault) activation model}, describing the prerequisites for fault activation;
\item The \textbf{error model}, describing detectable error states in the investigated system;
\item The \textbf{failure model}, describing different ways of failure, i.e., externally visible deviations from the specified system behaviour.
\end{itemize}

The differentiation of these models is a fundamental characteristic of our categorization scheme, as described later in Section \ref{tagnames}.

\section{Problem Statement and Research Questions}\label{rqs}
% TODO: rewrite as real questions+problems
The overall goal of this meta-study is to structure existing research on software failure cause models.
Such a structuring of existing models is beneficial in multiple ways:
For researchers, we want to provide insights into current trends and research gaps, to guide further research. Furthermore, as our meta-study is based on a systematic and exhaustive search of major databases, we ultimately want to enable researchers to look up publications of their interest based on a semantic structuring.  
For practitioners, theoretical work needs to be made more accessible, hence tag categories (discussed in Section \ref{tagnames}) should refer to practical software engineering aspects. 

To structure the broad research field of software failure causes, we aimed to answer the following research questions in our \gls{sms}:

\begin{researchquestion}\label{whatclasses}
What kind of models -- of faults, errors, failures, or fault activations -- are studied and referenced by the scientific community?
\end{researchquestion}

The terminology described in Section \ref{terminology} and in \cite{terminology} provides a starting point to answer this question. Based on the different steps or preconditions towards a software failure, we strive to understand how much and which research effort is focussed on each step.
This question is answered by semantically analyzing the abstracts of all relevant papers (manually and independently by two of the authors) and categorizing it into the four classes mentioned above.

\begin{researchquestion}\label{whatterminology}
What are the most common terminology models for describing software problems?
\end{researchquestion}

As discussed above in Section \ref{terminology}, we observed that software dependability research uses various divergent sets of terminology. One goal of our study is to verify this observation by systematically examining the terminology used in each publication and tagging each distinct set of terminology we came across differently. We thus hope to find out which terminology is most popular in different domains of software dependability research.

\begin{researchquestion}\label{howcategorized}
How are software problems categorized by related work?
\end{researchquestion}

One way of categorizing software failure cause models -- as fault, error, fault activation or failure model -- has already been discussed. We are interested in which other groupings and dimensions are relevant to precisely classify a software problem. To answer this question, we develop eight tag categories based on the scanned search result papers. 

\begin{researchquestion}\label{whattrends}
What are the focal points in research on software failure causes? Do research gaps exist?
\end{researchquestion}

Research on software failure causes is diverse, unstructured and hardly accessible. Once an understanding of which relevant categories of software failure causes exist is established, the distribution of research papers across these categories becomes interesting. We investigate whether certain categories, or combinations thereof, exhibit eye-catching clusters (trending research topics) or sparsity (either compelling research gaps, or irrelevant areas).

\section{Research Methodology: Systematic Mapping Study}\label{sms}

This section briefly describes the systematic mapping process, a research methodology we applied in the presented work. A more detailed explanation can be found in the article by Petersen et al. \cite{petersen2008systematic}.

A \acrfull{sms} is a defined scientific method to \enquote{build a classification scheme and structure a (software engineering) field of interest}. Originating from the medical science domain, the systematic mapping approach has gained attention also in the science of software engineering, due to the growing need for structured, evidence based approaches. This approach is applicable when an area of research has matured to an extent that many diverse publications exist which need to be structured and summarized, in order to gain insights into emerging trends and patterns. A systematic map helps in surveying and analysing a broad research field of interest -- in our case, the entirety of papers presenting software failure causes.

The \gls{sms} comprises the following steps:
\begin{enumerate}
\item One or more \textbf{research questions}, reflecting the goal of the study, are defined.
\item A list of relevant databases and publication forums for answering the research questions is compiled. A primary search for relevant papers is conducted on these databases. The search should be carried out in a systematic way, using a well defined \textbf{search string} which includes all the sub-topics of interest. Finding an adequate search string is an iterative process, it should be repeated and fine tuned to avoid bias.
\item Additional \textbf{inclusion and exclusion criteria} are defined. These criteria serve the purpose of selecting only papers relevant for answering the research questions, and filtering out publications which do not satisfy certain formal criteria.
\item Based on search results and the inclusion and exclusion criteria, a final list of publications to be included in the mapping is assembled. For these papers, a \textbf{keyword classification / tagging of the abstracts} takes place. Each abstract is assigned a number of keywords or \textbf{tags} (the term we will use subsequently), which describe the publication's position and focus with regard to the research questions.
\item The tags finally form the basis of the \textbf{systematic map}, i.e., the data set containing the relevant tags and the amount of papers assigned to them. The systematic map can be visualized for instance as a bubble plot, where trends and hotspots can be identified. Subsequent to a \gls{sms}, the trending areas in the map can become topics for more detailed \emph{systematic literature reviews} \cite{keele2007guidelines}.
\end{enumerate}

\section{Design of the Study}\label{methodology}

As sources for literature included in the mapping, we searched major online databases and publication platforms used by the software reliability and software engineering community, as shown in Table \ref{tab:searchstrings} (appendix). We omitted the ResearchGate database, as its focus lies more upon social networking aspects and its search features are not powerful enough for our purpose. We also omitted all public databases where English is not the main language.

Initially, we intended to use one generic search string for all databases, in order to achieve the best possible comparability of results. This turned out to be impossible in practice. Since the databases we searched use different query syntax formats of varying expressiveness, we had to express our search objectives in a tailored way for each of them.

Due to the diverging terminologies representing failure causes, we had to cover all combinations of the form \\ \texttt{\{fault|error|defect|bug\}} \texttt{+} \texttt{\{model|classification|taxonomy\}} \\ in the queries. The number of false positives in our search results was therefore large. Especially the term \texttt{classification} attracted many papers from the machine learning and data science domains. This was hard to filter out by tailored queries, since classification is a valid term in software engineering research too. We ended up applying a manual filtering process here.

Furthermore, we manually added publications we deemed relevant. The manual adding of literature besides using a search string is not unusual, it was demonstrated for example by Walia et al. \cite{dogan_web_2014} in their systematic literature review.

We used the group feature of the \emph{Zotero} bibliography management tool\footnote{\url{https://www.zotero.org/}, \today} for the purpose of maintaining a collaborative bibliography collection.

\subsection{Inclusion and Exclusion Criteria}\label{inclusion_criteria}

The \textbf{inclusion criteria} are intended to put focus on the defined research questions. Each publication in our study, therefore, meets at least one of the following criteria:

\begin{itemize}
	\item Presentation of an own classification, model or taxonomy of software failures or failure causes. The paper abstract indicates that an own software failure cause model might be presented.
    \item Extension or critique of an existing classification or model of software failures or failure causes.
    \item A well-known meta study of software failures or failure causes is referenced and extended or discussed.
\end{itemize}

The \textbf{exclusion criteria}, which lead to the filtering out of a paper, were the following:

\begin{itemize}
    \item Focus mainly on hardware or hardware description languages.
    \item Research which is not available in English language.
    \item Research which is not published in conference, workshop or journal papers, or as an indexed technical report in one of the searched databases.
\end{itemize}

The consistent definition of inclusion/exclusion criteria goes hand in hand with the formulation of the search string. Both have to be validated in combination. When starting the study, we had a set of relevant publications in mind which were known to us. Thus, a first validity test for the search string and the inclusion and exclusion criteria was whether these publications appeared in the search results. When, in the process of the mapping, through related work, we identified further \enquote{very relevant} papers, we included them in our set of \enquote{must-have} publications, and re-iterated the search string accordingly.

\section{Categorization of Software Failure Cause Models}\label{tagnames}
%TODO: try to present this as contribution

When discussing a research field as broad as that of software dependability, establishing a common structure is essential.
In this section, we contribute a categorization of software failure cause models. 

Based on the scanned papers, our background knowledge, and practical experience, we therefore defined eight categories which represent relevant aspects for characterizing software failure cause models.
For each category, a number of tags describing the sub-categories within was defined.
This categorization is the basis for the evaluation of the study's results, which is presented in Section \ref{results}.

\subsection{Tag Categories for Software Failure Cause Models}

\begin{figure*}[ht!]
\centering
\includegraphics[width=0.8\linewidth]{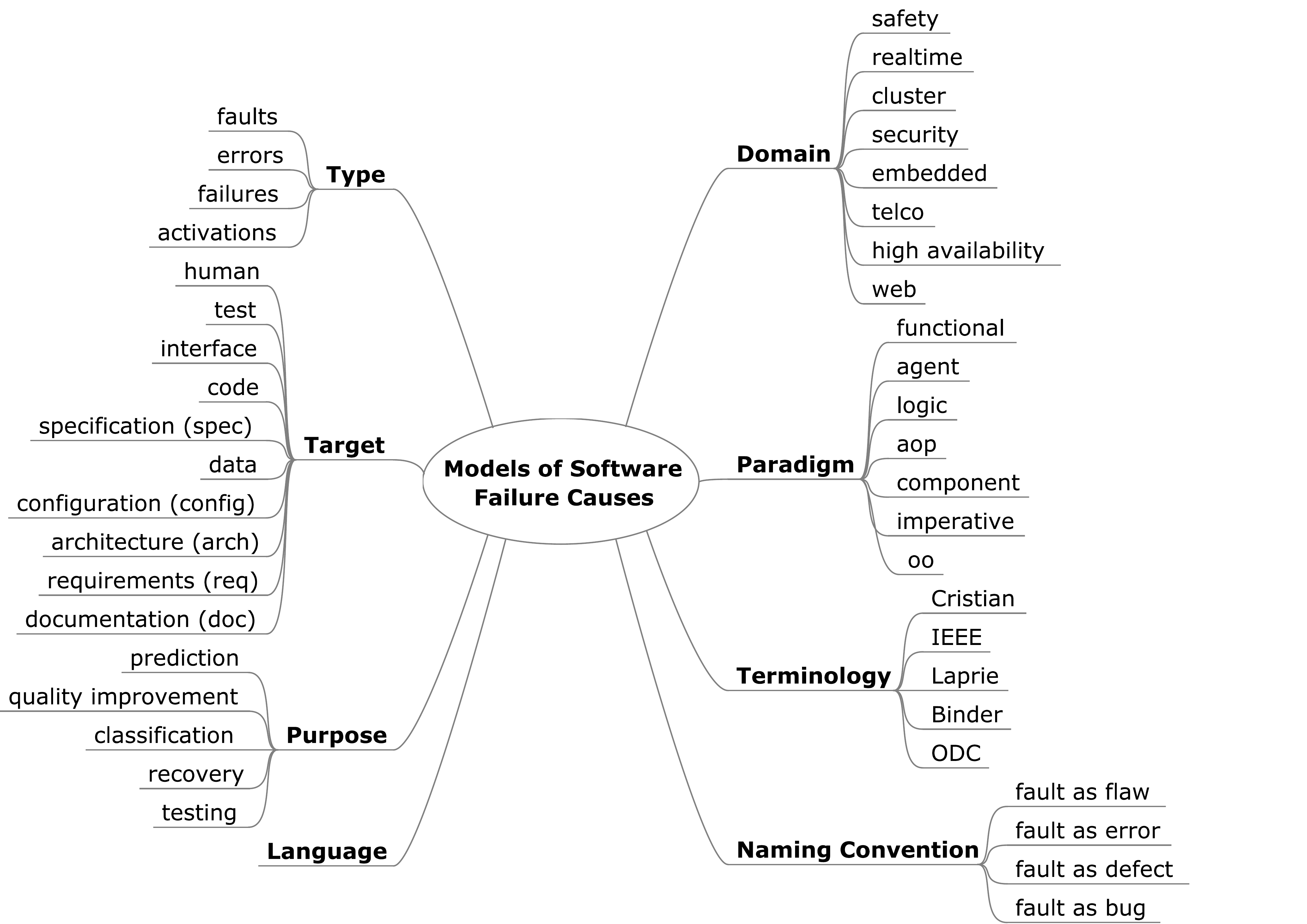}
\caption[Categorization of tag names used in the mapping]{Categorization of tag names used in the mapping. We used eight independent categories of tags, which were established prior to the tagging, after a preliminary scan of the relevant papers. The tag categories are orthogonal to each other.}
\label{fig:tags}
\end{figure*}

Figure \ref{fig:tags} shows our the categorization of tags, which was used to structure and analyze the research papers.
%Table \ref{tab:modeltypes} describes the tag categories in more detail.

The \textbf{Type of Model} category is pivotal to our study, as we use it to apply our terminology foundation to classify each paper with the same (Laprie/\avz-based) vocabulary. We put special emphasis on this tag category, because it involves a deeper understanding of the paper than the other tags, and tagging includes our own interpretation of the contribution of the individual papers.
According to what the model describes, the following tags exist:
\begin{itemize}
	\item \texttt{\textbf{fault}} -- causes of error states in a program, usually mistakes in the program code.
	\item \texttt{\textbf{activations}} -- conditions on the program and environment state, which must hold so a fault is activated and becomes an error. %Papers in this category:{\footnotesize\input{../data/citations/references_model_of_activations}}
	\item \texttt{\textbf{errors}} -- undesired states of the program, which may lead to a failure. %Papers in this category:{\footnotesize\input{../data/citations/references_model_of_errors}}
	\item \texttt{\textbf{failures}} -- externally visible deviations from the program's specification. %Papers in this category:{\footnotesize\input{../data/citations/references_model_of_failures}}
	\item \texttt{\textbf{meta}} -- description of methodology for obtaining a fault, activation, error or failure model from software. %Papers in this category:{\footnotesize\input{../data/citations/references_meta}}
\end{itemize}
The \textbf{Target} category describes what aspect or software artefact is targeted by the model. Within this category, we use the following tags:
\begin{itemize}
	\item \texttt{\textbf{code}} -- (static) source code, which can be compiled or interpreted and is thus executable.
	\item \texttt{\textbf{interface}} -- interface(s) between software different components.
	\item \texttt{\textbf{documentation}} -- non-executable source code documentation artefacts.
	\item \texttt{\textbf{requirements}} -- assets from the requirements elicitation phase of a software development process.
	\item \texttt{\textbf{specification}} -- non-executable description (separate from the source code) of what the desired behaviour is\footnote{We are well aware that a debate about the precise boundary between \enquote{program} and \enquote{specification} exists. We use the term \enquote{specification} for behaviour descriptions which can be formal and analysable, but are not intended for actual execution.}
	\item \texttt{\textbf{architecture}} -- the higher level structure of software elements and their interaction patterns. 
	\item \texttt{\textbf{test}} -- code or procedures for testing the system functionality.
	\item \texttt{\textbf{configuration}} -- documents and scripts for deployment, compilation or runtime property configuration.
	\item \texttt{\textbf{human}} -- human behaviour when interacting with the system, also including usability aspects.
\end{itemize}

The \textbf{Purpose} category expresses the purpose of the proposed research.
\begin{itemize}
	\item \texttt{\textbf{testing}} -- derive better test cases or testing strategies, improve test coverage.
	\item \texttt{\textbf{prediction}} -- predict the occurrence of further failures, errors, or defect findings.
	\item \texttt{\textbf{classification}} -- categorize and classify faults, errors, and defects.
	\item \texttt{\textbf{quality improvement}} -- find process modifications, root cause analyses or other means to improve overall software quality.
	\item \texttt{\textbf{recovery}} -- derive automatic recovery and/or restoration mechanisms in the software.
\end{itemize}

The \textbf{Domain} category describes the main application domain of the model. Such domains include among others \texttt{safety}(-critical), \texttt{realtime} and \texttt{embedded}.
The \textbf{Language} category is used to tag papers which mainly target a specific programming language.
The \textbf{Paradigm} category characterizes the programming paradigm, if the paper focusses on a specific one. For instance, we encountered papers providing a conceptualization of software failure causes, which deals primarily with the \texttt{imperative}, \texttt{oo} or \texttt{logic} programming paradigms.

Finally, we use two tag categories for describing the original terminology used: Tags from the \textbf{Naming} category answer the question \enquote{If a terminology diverging from the Laprie/\avz terminology is used, how are faults named in the paper?}.
We observed that various terms which are used synonymously for \enquote{fault} exist: \texttt{flaw}, \texttt{error}, \texttt{defect} and \texttt{bug}.
Tags from the \textbf{Terminology} category are used to tag papers which are explicitly based upon previous failure cause terminologies.
We take the following base terminologies from literature into account:

Traditionally, the Laprie/\avz terminology \cite{avizienis2004basic}, which we also adopt, has been used for describing both software and hardware systems. Cristian \cite{cristian} described a software fault model, which focusses on distributed systems, and has been used mainly in that domain. Orthogonal Defect Classification (ODC) \cite{chillarege1992orthogonal} is an integrated approach to classify software defects in the context of the entire development process.
The IEEE software engineering glossary \cite{radatz1990ieee}, as already mentioned, also defines some vocabulary for describing software failure causes.
Binder \cite{binder2000testing} published a book on software testing which is includes model of software faults.

%\input{tagdescription.tex}
% We evaluate and discuss the tagging process in Section \ref{discuss:tag_categorization}.
\subsection{Consensus on Tags}\label{tagconsensus}
%\todo{Implementierungsdetail? Staerken/Schwaechen? Wieviele Reviewer sollten einbezogen werden?}
To decrease the likelihood of mistakes and to increase the independence of opinions, the two involved reviewers tagged all papers separately and without discussing them together. After this first round of individual tagging, consensus on the tags was ensured in a joint session.

The \enquote{Type of Model} category for clustering the failure cause models is most pivotal and gives the strongest hint about the focus of the paper, as it describes which kind of failure cause model is discussed. This category directly comes from our terminology foundation (see Section \ref{terminology}). Tagging papers in this category includes a semantic interpretation of the paper content. Therefore, the goal of the joint session was to agree upon the same set of \enquote{Type of Model} tags. Initially, these sets were different for 73 papers, but not completely disjunct. For example, one reviewer had tagged \enquote{model of faults} and the other \enquote{model of faults, model of activations} for the same paper. The joint tagging session helped to eliminate some accidental mistakes and also sharpened the distinction between the different tags.

Papers discussing \emph{usability problems} were tagged as \enquote{model of failures}, because usability becomes visible externally to the user, and is thus a deviation from the (implicit) system specification.
Papers discussing \emph{distributed fault injection} were tagged as presenting a \enquote{model of faults}, because the failure of a sub-component becomes a fault, i.e., a cause for an error state, in the investigated distributed system. This corresponds to Laprie/\avz's concept of error propagation chain \cite{avizienis2004basic}.
Papers presenting \emph{reliability growth models} were tagged as \enquote{model of faults}, because they mathematically model the occurrence of bugs, which are faults in the source code.
Papers with a strong \emph{security focus} were hard to categorize, because they often intermingled error states and fault patterns -- both are relevant for discussing potential exploits and attacks. Therefore, when in doubt, such papers were tagged \enquote{model of faults, model of errors}.

%At the end of the joint tagging session, having re-visited many critical papers, both reviewers gained confidence in having the same understanding of what each paper aimed at communicating.
Apart from the \enquote{Type of Model} category, the other categories did not require much discussion. There, the sets of tags, when not identical, were merged to result in the union of both sets. Commonly, one set of tags was a superset of the other, so by merging the tags, we included the maximum of information.

%\clearpage
\section{Results}\label{results}

This section presents the \gls{sms} results, concentrating on certain points of interest\footnote{For the complete dataset of citations and tags, refer to \texttt{\url{https://gist.github.com/laena/9b514aa89cc0f690a367/download}}.} In total, both reviewers provided tags for 156 publications.

\begin{figure}[h!]
\includegraphics[width=\linewidth]{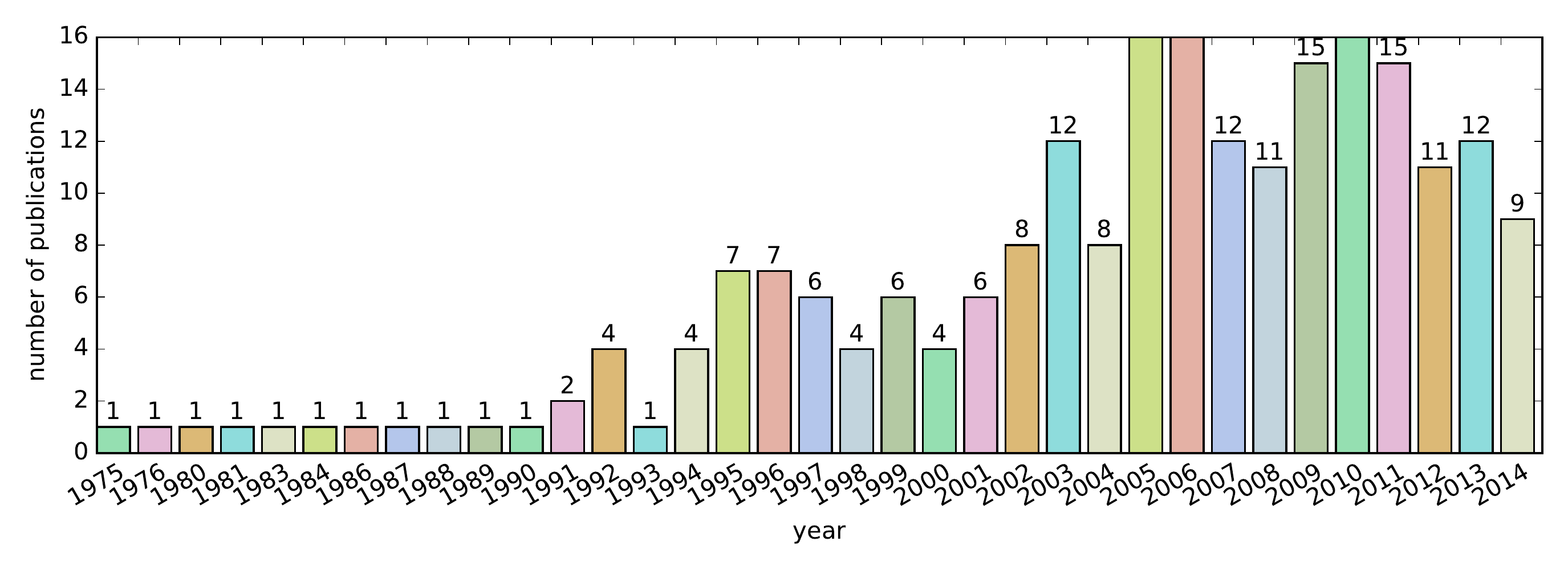}
\caption{Number of papers in our study by publishing year.}
\label{hist_years}
\end{figure}

As shown in Figure \ref{hist_years}, the publishing years of the included publications span several decades, from 1975 to today. There has been a recent trend towards more publications in the area. The recent decline of the number of publications, especially in 2014, might be explained by the fact that not all search indices have been updated to include the latest publications at the time of search.

Table \ref{totalfacetcount} shows how many of these 156 publications were tagged with one or multiple tags per category.

\begin{table}[h!]
  \begin{tabularx}{\linewidth}{lXX}
  \toprule
  All papers & 156 & 100\% \\
  \midrule
  Type & 148 & 94.87\% \\
  Convention & 68 & 43.59\% \\
  Terminology Basis & 43 & 27.56\% \\
  Target & 124 & 79.49\% \\
  Domain & 62  & 44.23\% \\
  Purpose & 150 & 96.15\% \\
  Language & 28 & 17.95\% \\
  \bottomrule
  \end{tabularx}
  \caption{Total number and percentage of papers tagged per category.}
  \label{totalfacetcount}
\end{table}

\begin{figure*}[ht!]
\centering
\includegraphics[width=\linewidth]{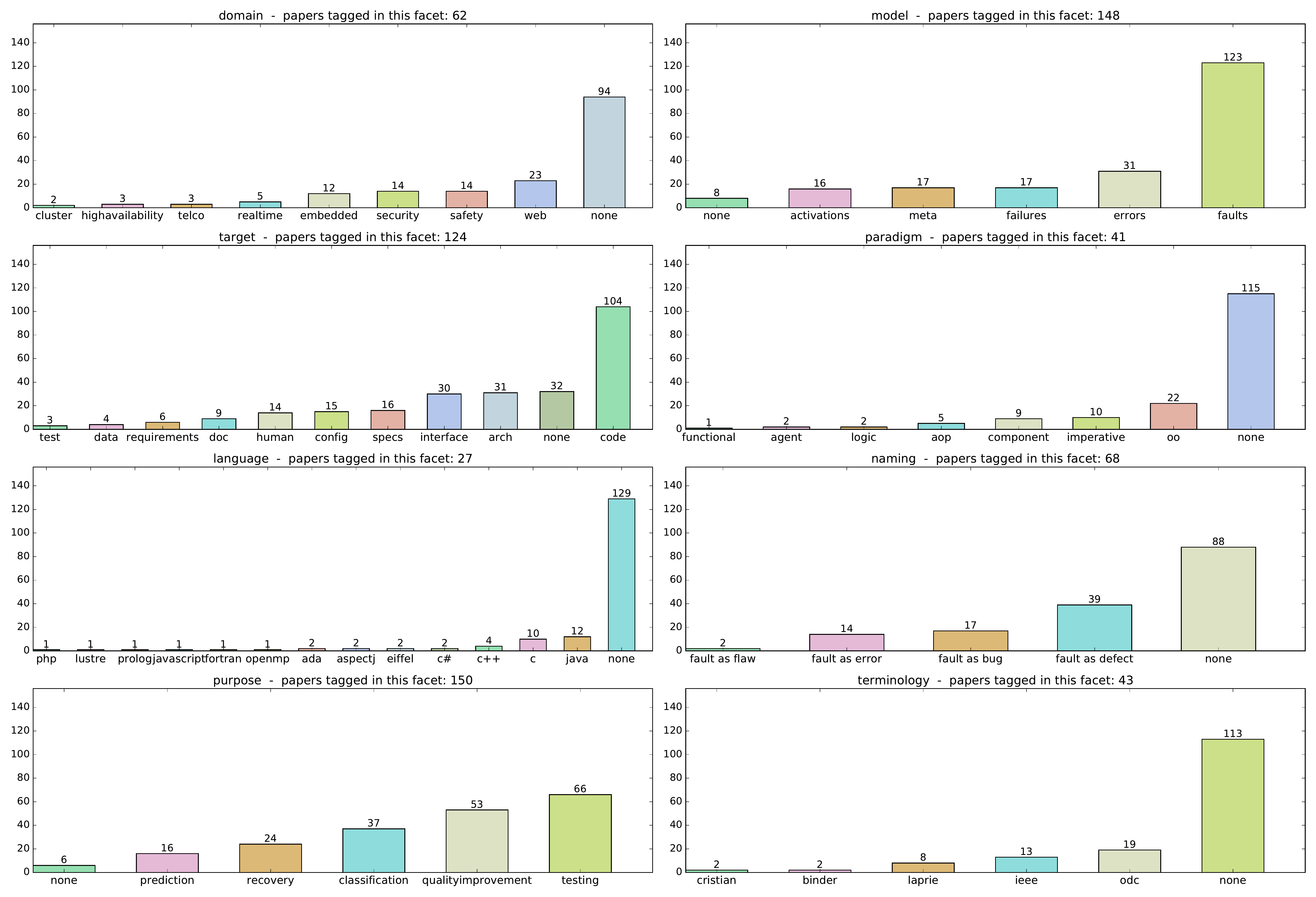}
\caption{Tag occurrences within each category. The y axis shows the number of tag occurrences (from a total of 156 papers).}
\label{hist_all_tags}
\end{figure*}

It has to be noted that the dataset does not strictly consist of nominal or categorical data: within each category an arbitrary number of tags can be applied to one paper, i.e., tags are not mutually exclusive within a category. Therefore, in all subsequent statistics, the total tag count per category not necessarily sums up to the total number of papers.
%\todo{Mehr Statistiken? Least cited, most cited, Popular authors, publishing platforms}

As mentioned in Section \ref{tagconsensus}, we ensured that each paper was tagged with one or multiple tags describing the type of model, which the reviewers both consented to.
Table \ref{typereferences} shows the references per tag in the \enquote{Type of Model} category.

Figure \ref{hist_all_tags} depicts number of tag occurrences for all tag categories.

\begin{table}[h!]
  \begin{tabularx}{\linewidth}{lXX}
  \toprule
  Model of... & References  \\
  \midrule
  faults & {\tiny\cite{aichernig_mutation_2009,aissa_quantifying_2010,alexander_syntactic_2002,ammar_fault_2001,arisholm_systematic_2010,aslam_taxonomy_1995,avizienis_approaches_1976,babu_fault_2009,bartel_model_2011,basili_comparing_1987,belli_towards_1996,bellman_modeling_1990,black_counting_2011,bougie_comparative_2010,bowen_standard_1980,bozzano_safety_2011,bruning_fault_2007,bruntink_discovering_2006,bsekken_candidate_2006,buchler_security_2013,can_eliminating_2007,chang_defect_2007,ciupa_finding_2008,ciupa_number_2011,claesson_xbw_1998,collofello_proposed_1984,damm_identification_2005,de_lemos_architecting_2009,dogan_web_2014,drebes_kernel-based_2006,du_bousquet_mutation_2007,du_testing_2000,dumitras_why_2009,eisenstadt_my_1997,el_emam_repeatability_1998,elbaum_software_1999,en-nouaary_fault_1999,endres_analysis_1975,evanco_architectural_2003,farchi_concurrent_2003,ferrari_characterising_2010,freimut_industrial_2005,furuyama_analysis_1997,gao_out--bounds_2007,gario_fail-safe_2014,gong_expressions_2003,hanchate_defect_2010,harrold_approach_1997,hayes_building_2003,hayes_case_2006,he_general_2001,hecht_proposal_1995,holling_fault_2014,huang_autoodc:_2011,hui_taxonomy_2010,hunny_osdc:_2013,jacques-silva_network-level_2006,joshi_proposal_2005,jung_layered_2011,kandasamy_transparent_2003,karthik_defect_2010,khomh_exploratory_2012,kidwell_decision_2011,kim_soar:_2009,ko_development_2003,kumar_study_2009,landwehr_taxonomy_1994,laski_programming_1997,leszak_case_2000,leung_selective_1995,li_classification_2010,li_design_2002,li_mining_2010,li_policy-based_2011,liu_software_2011,looker_assessing_2003,looker_assessing_2003-1,looker_ontology-based_2005,looker_pedagogic_2005,lopes_application_2013,luo_control-flow_1992,mantyla_what_2009,marchetto_empirical_2007,mariani_fault_2003,mellegard_light-weight_2012,munson_toward_2002,nadeem_automated_2006,nagappan_preliminary_2004,nagwani_comparative_2014,nagwani_generating_2013,nakashima_analysis_1999,nakata_fault_2011,nath_improvement_2012,nigam_classifying_2012,offutt_fault_2001,oppenheimer_studying_2002,oyetoyan_comparison_2013,padgham_model-based_2013,pan_toward_2009,paradkar_quest_2006,paradkar_towards_2004,raja_all_2013,ricca_web_2005,robillard_saying_2007,roychowdhury_algorithm-based_1996,runeson_what_2006,silva_practical_1998,straub_edf:_1991,stringfellow_empirical_2002,sullivan_comparison_1992,tewary_fault_1994,thung_automatic_2012,torchiano_are_2011,trivedi_recovery_2011,tsipenyuk_seven_2005,vaidyanathan_comprehensive_2005,vilbergsdottir_assessing_2014,vilbergsdottir_classification_2006,weber_software_2005,wei_model_2011,xu_adam:_2012,xu_prioritizing_2010,yang_path_1992,yen_efficiently_1995,yousef_analysis_2009,yu_analysis_1988,zaryabi_neural_2014,zhang_research_2010,zheng_value_2006,zhou_rule_2009}} \\
  errors & {\tiny\cite{aissa_quantifying_2010,anderson_framework_1983,avizienis_approaches_1976,aysan_error_2008,black_counting_2011,borin_software-based_2006,bozzano_safety_2011,bruning_fault_2007,darmaillacq_security_2008,duran_proposal_2007,eisenstadt_my_1997,gamrad_formalization_2008,gong_generating_2008,hecht_proposal_1995,ignat_soft-error_2006,lackovic_taxonomy_2010,landwehr_taxonomy_1994,laski_programming_1997,leszak_case_2000,li_policy-based_2011,liu_software_2011,murthy_quality_2011,owens_software_1996,ricca_web_2005,straub_edf:_1991,tan_hierarchical_2010,trivedi_recovery_2011,tsipenyuk_seven_2005,weber_software_2005,wei_model_2011,xu_adam:_2012,yang_path_1992}} \\
  activations & {\tiny\cite{aysan_error_2008,ben-asher_noise_2006,bozzano_safety_2011,da_silva_design_2009,darmaillacq_security_2008,du_testing_2000,gamrad_formalization_2008,gario_fail-safe_2014,godskesen_fault_1999,he_general_2001,hecht_proposal_1995,laski_programming_1997,leszak_case_2000,li_design_2002,liu_software_2011,pan_dimensionality_1999,sullivan_comparison_1992,tsipenyuk_seven_2005}} \\
  failures & {\tiny\cite{hecht_proposal_1995,huynh_another_2009,jin_markov_2012,joshi_proposal_2005,kandasamy_transparent_2003,li_policy-based_2011,looker_pedagogic_2005,ma_web_2007,nesterenko_dining_2002,padgham_model-based_2013,pan_dimensionality_1999,raz_research_2002,ricca_web_2005,silva_practical_1998,straub_edf:_1991,thakur_analyze-now-environment_1996,thelin_evaluation_2004,vilbergsdottir_assessing_2014,vilbergsdottir_classification_2006,widder_synchronous_2007}} \\
  meta & {\tiny\cite{belli_towards_1996,claesson_xbw_1998,dogan_web_2014,el_emam_repeatability_1998,freimut_industrial_2005,gras_end--end_2004,he_general_2001,hecht_proposal_1995,holling_fault_2014,kim_soar:_2009,ko_development_2003,looker_pedagogic_2005,nagwani_generating_2013,paradkar_salt-integrated_2000,paradkar_towards_2004,wagner_defect_2008,zhou_rule_2009}} \\
  \bottomrule
  \end{tabularx}
  \caption{References for tag category \enquote{Type of Model}.}
  \label{typereferences}
\end{table}

% \subsection{Visualization of Results}\label{visualization}
% Traditionally, one way of visualizing the results of a \gls{sms} is using \emph{two-dimensional bubble charts} to highlight clusters of research (represented by the bubble size) within a 2D grid defined by two categories or tag categories. A bubble at coordinate $(x_i,y_j)$ -- corresponding to the joint usage of tag $x_i$ from category $X$ and tag $y_j$ from category $Y$ -- is usually scaled according to the total number of occurrences of this tag combination.

% In our case, the number of papers tagged varied strongly for each category. Therefore, tag combinations where one tag comes from a \enquote{popular} category tend to be visualized with bigger bubbles and thus over-emphasized, if the interest lies upon the mutual dependence of two categories. In other words, consider an extreme case: even if all tags from a not frequently used category correlated strongly with another tag, the bubbles might seem small due to the low absolute number of tags from that category. Thus, interesting correspondences between tags might be overseen.

In the following plots, we therefore visualize tag combinations using a \emph{correspondence value}. The correspondence value is our measure of correlation between two tags. Consider a set of papers tagged simultaneously with tags $A$ and $B$, $T_{A \wedge B}$, and the set of papers $T_A$ and $T_B$ tagged with $A$ and $B$ respectively. The correspondence value is then computed as follows:
\[
Correspondence = \frac{\abs{T_{A \wedge B}}}{\abs{T_A} + \abs{T_B}} \times 2
\]
We scale the value by multiplying it with the number of dimensions considered, so that the maximum value becomes $1.0$.
We thus have a relative measure of how frequently two tags occur together, ranging from $0.0$ to $1.0$. If all papers tagged with $A$ were also tagged with $B$ and vice versa, this is the highest possible correspondence between two tags, with a resulting value of $1.0$. We also provide the absolute occurrence count and its relative frequency with regard to all papers in the bubble plots.

\begin{figure}[h!]
\centering
\includegraphics[width=.7\linewidth]{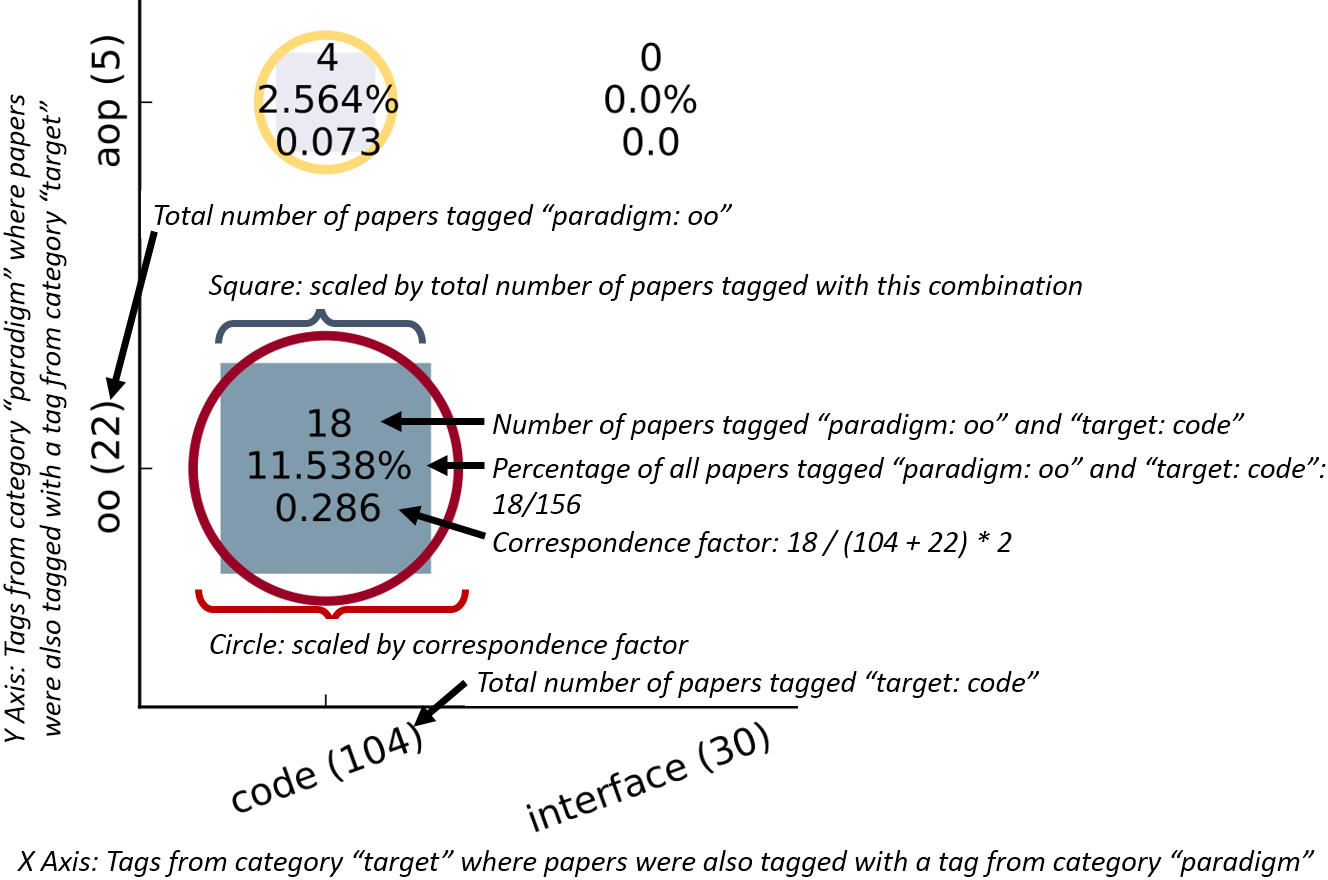}
\caption{Illustration of the visualization schema. For each tag combination, the absolute count and the correspondence value are shown. At the axes, tag names are annotated with the total number of occurrences of that tag in all papers.}
\label{fig:visualization}
\end{figure}

In our 2D bubble visualizations, we use \emph{circles} and \emph{squares} as visualization primitives.
As an example, consider Figure \ref{fig:visualization}, which shows the entry at (\texttt{paradigm: oo}, \texttt{target: code}). The large circle is scaled with the correspondence value describing that \emph{a large fraction of the papers tagged \enquote{oo} or \enquote{code} was tagged with both these tags}. The large square indicates that \emph{a large absolute number of papers was tagged with both tags}. The absolute and correspondence values are two different views on the data, depending on whether the focus is on trending research areas with a large number of publications, or on the correspondence and interdependence between two individual tags.

For the sake of understandability, we omit rows and columns where all tag combinations have occurred zero times. This means that tags from one dimension, which have never been used together with any tag from the other dimension, are not visualized. We justify this omission by arguing that our sample size is too small to draw meaningful conclusions for tag combinations which occur very rarely.

\begin{figure}[h!]
\centerline{\includegraphics[width=.85\textwidth,height=.85\textheight,keepaspectratio]{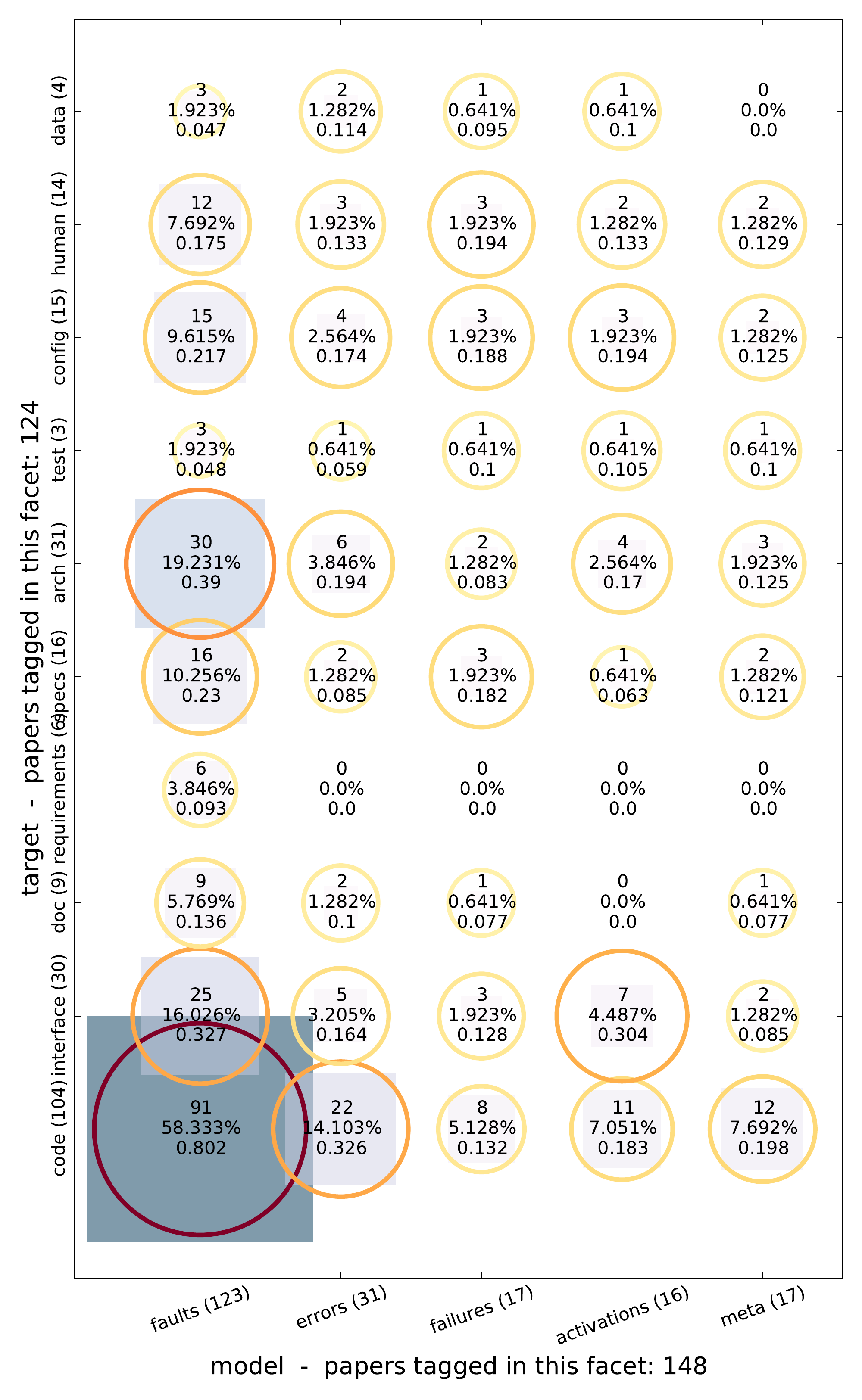}}
\caption{Occurrences of tag combinations from the model and target categories. This figure shows that a large number of papers is concerned with fault models which primarily target code (58\% of all papers). Models of errors dealing with code are also seen frequently, with 22 such papers in our study. It is also interesting to note that if a model of activations is discussed, as in 16 out of 156 papers, this model frequently targets the interface. \interpretation{The fact that the combination \enquote{model of faults} and \enquote{target is code} occurs frequently justifies our decision to interpret faults in software as design/code imperfections (see Section \ref{terminology}). The fact that we excluded papers with a focus on hardware could explain the rarity of error models targeting data -- data corruption is frequently discussed in hardware fault and error models.}}
\label{scatter_target_model}
\end{figure}

\begin{figure}[h!]
\centerline{\includegraphics[width=.85\textwidth,height=.85\textheight,keepaspectratio]{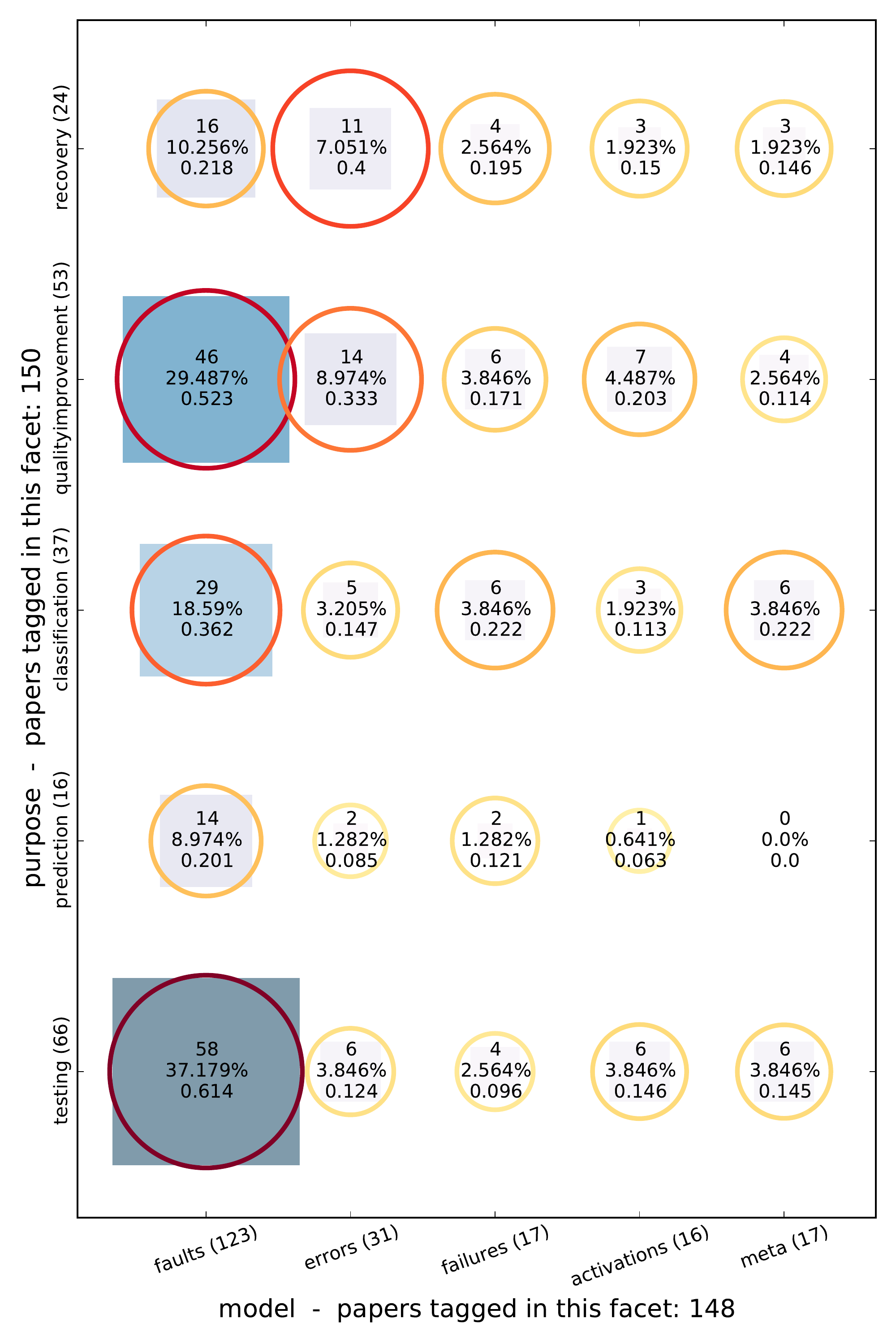}}
\caption{Occurrences of tag combinations from the model and purpose categories. Not very surprisingly, fault models are mostly used for testing, quality improvement and classification, whereas error models are frequently used for error recovery and quality improvement. \interpretation{Testing not only assumes a fault model, but also a certain kind of erroneous behaviour which the tests try to detect. However, not much literature deals with testing and error models at the same time. Here, we see a need for future research.}}
\label{scatter_model_purpose}
\end{figure}

\begin{figure}[h!]
\centerline{\includegraphics[width=.85\textwidth,height=.85\textheight,keepaspectratio]{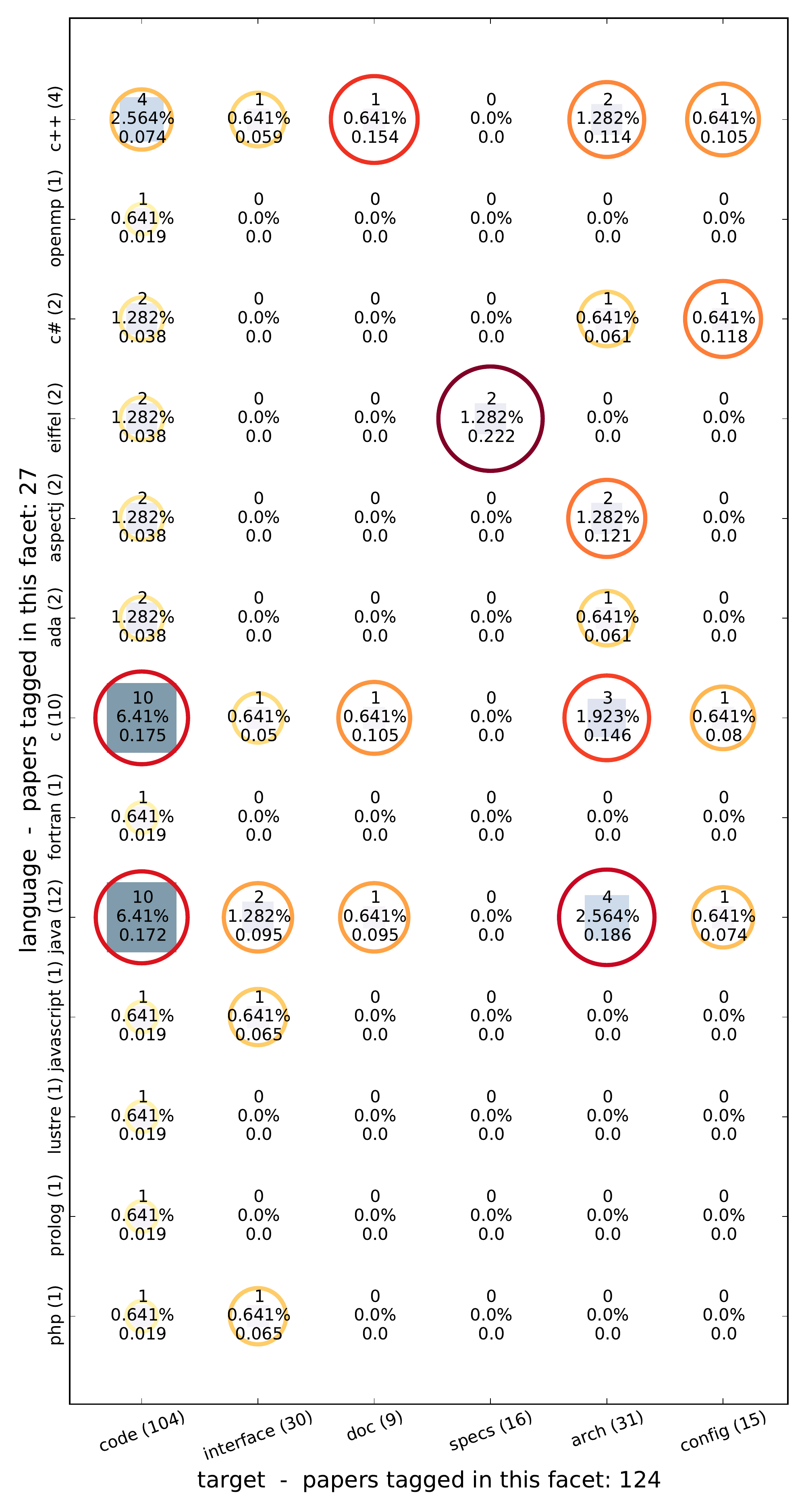}}
\caption{Occurrences of tag combinations from the target and language categories. All papers discussing Eiffel target the specification in their model. Papers discussing the languages Java and C, on the other hand, are more focussed on code and architecture. No language-specific requirements, human or data aspects obviously have been studied in the investigated set of papers. \interpretation{This visualization confirms the validity of the mapped data, as the results correspond to widely accepted facts about programming languages. For example, Eiffel is well-known for embedding specification within the language. The code focus in Java and C papers might be due to the fact that a large amount of bug data exists for these \enquote{traditional} programming languages.}}
\label{scatter_target_language}
\end{figure}

\begin{figure}[h!]
\centerline{\includegraphics[width=.85\textwidth,height=.85\textheight,keepaspectratio]{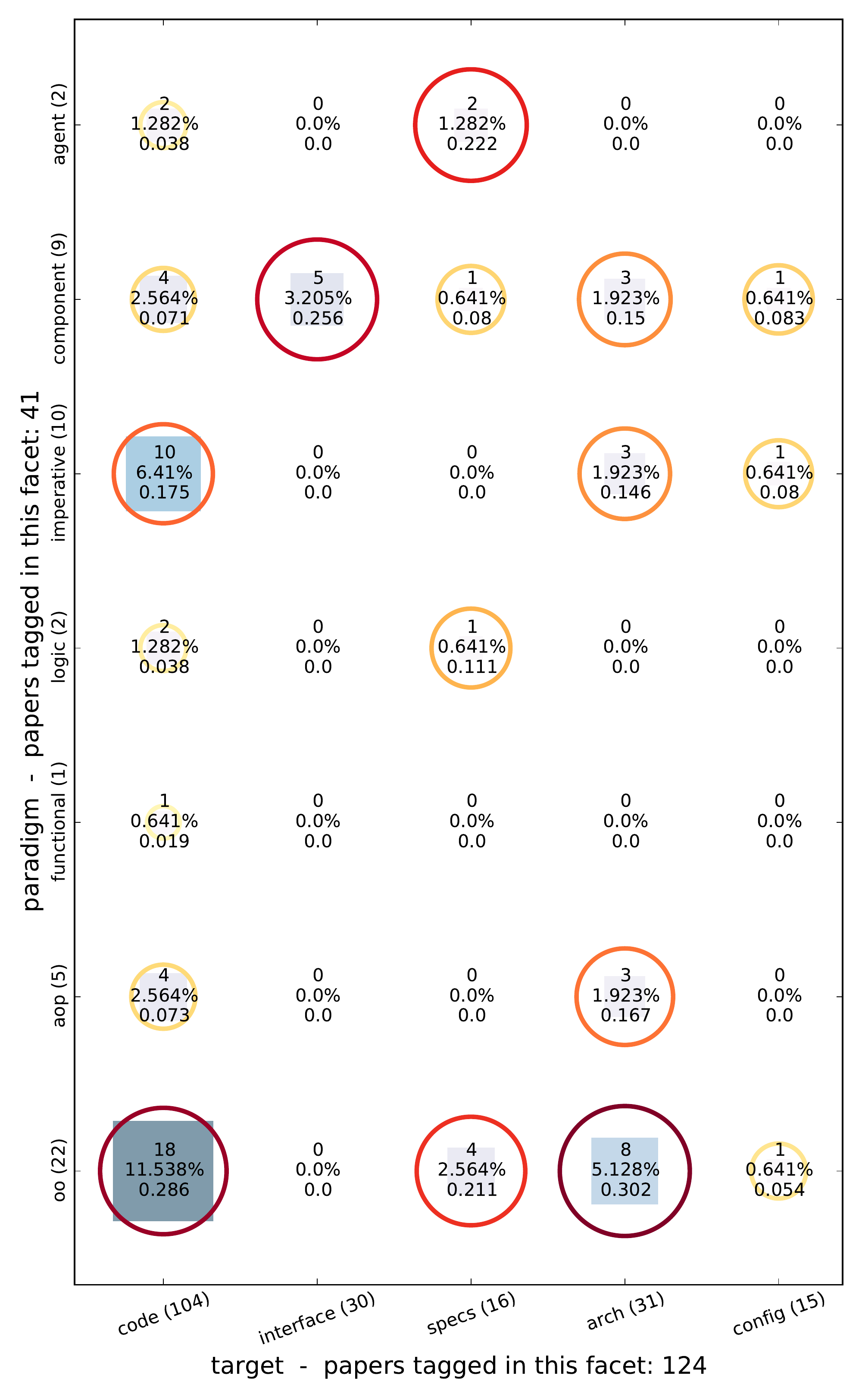}}
\caption{Occurrences of tag combinations from the target and paradigm categories. Architecture and code are mainly targeted in papers about object oriented software, whereas papers dealing with component based software focus more on interface problems. It is also interesting to note that specification is targeted mainly by agent-based and object oriented languages. \interpretation{Interaction is a major aspect of many software problems. Intuitively, interaction in component based software is defined by the interfaces. The object oriented community, on the other hand, focusses more on architectural issues as the root cause for software defects.}}
\label{scatter_target_paradigm}
\end{figure}

\begin{figure}[h!]
\centerline{\includegraphics[width=.85\textwidth,height=.85\textheight,keepaspectratio]{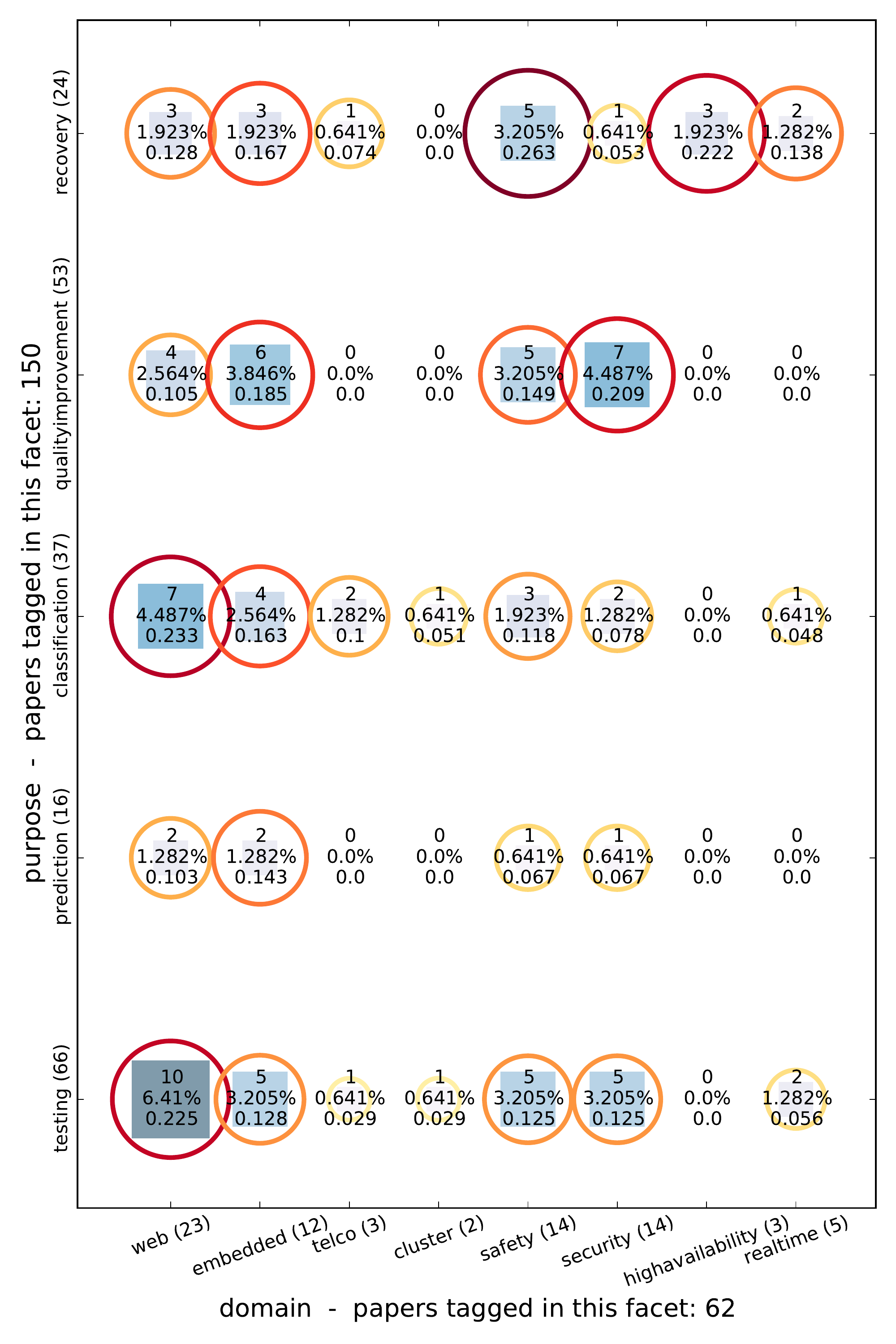}}
\caption{Occurrences of tag combinations from the domain and purpose categories. Papers dealing with web services frequently develop models for testing and the classification of failure causes. Recovery from errors tends to be studied primarily in safety-critical and high availability scenarios. \interpretation{The thorough discussion of error recovery schemes seems worthwhile in scenarios, were error propagation and outage have severe consequences, such as high availability and safety-critical applications.}}
\label{scatter_domain_purpose}
\end{figure}

\begin{figure}[h!]
\centerline{\includegraphics[width=.85\textwidth,height=.85\textheight,keepaspectratio]{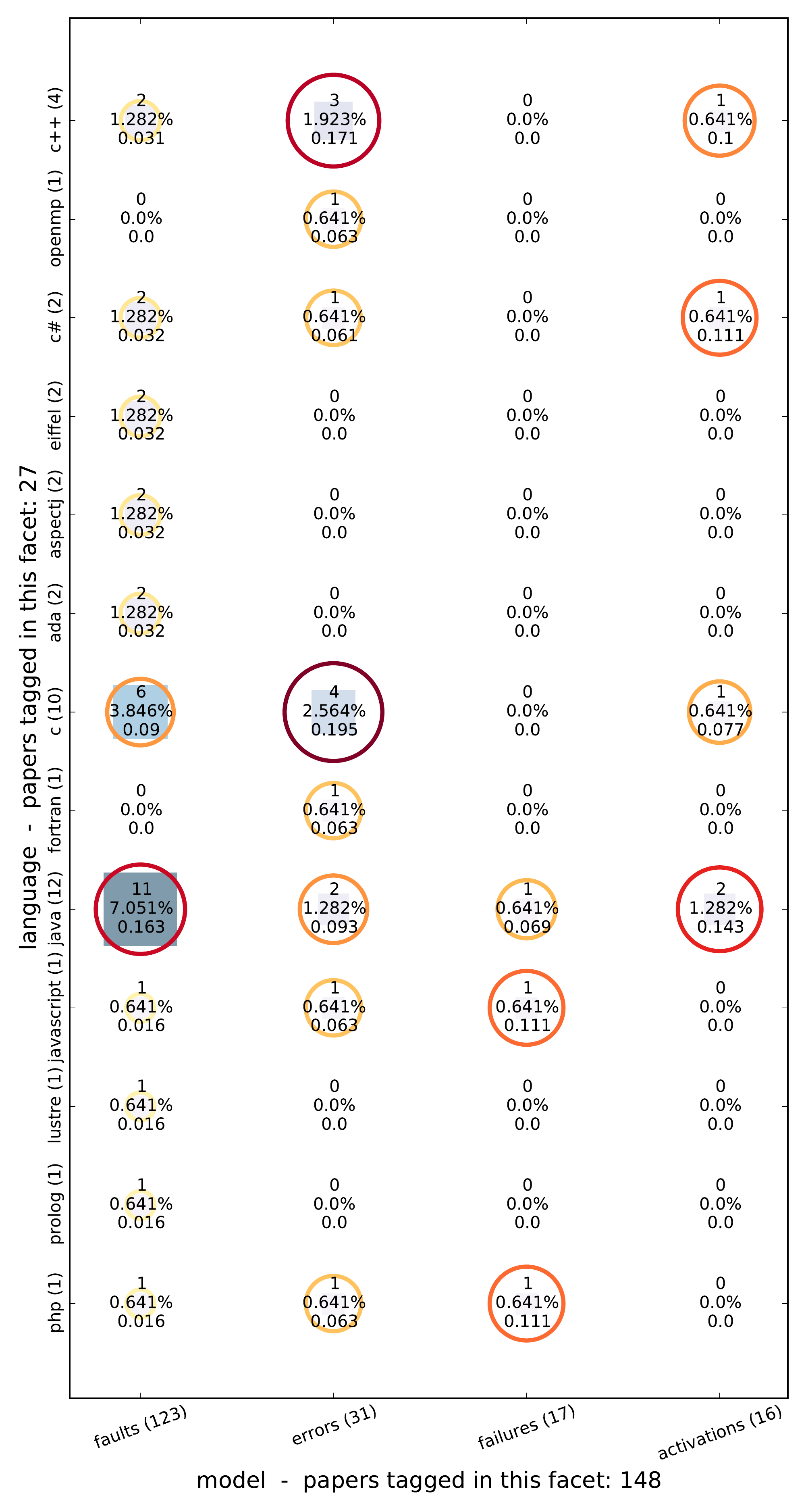}}
\caption{Occurrences of tag combinations from the model and language categories. Among the papers concerned with Java, code-based fault models are primarily studied. On the other hand, C and C++ papers tend to discuss error states. \interpretation{Due to the complexity of features such as pointer arithmetic, finding adequate fault patterns for C/C++ might be hard. In such languages without automated memory management, studying error states such as memory corruption might be more relevant than syntactic fault patterns.}}
\label{scatter_language_model}
\end{figure}

\begin{figure}[h!]
\centerline{\includegraphics[width=.85\textwidth,height=.85\textheight,keepaspectratio]{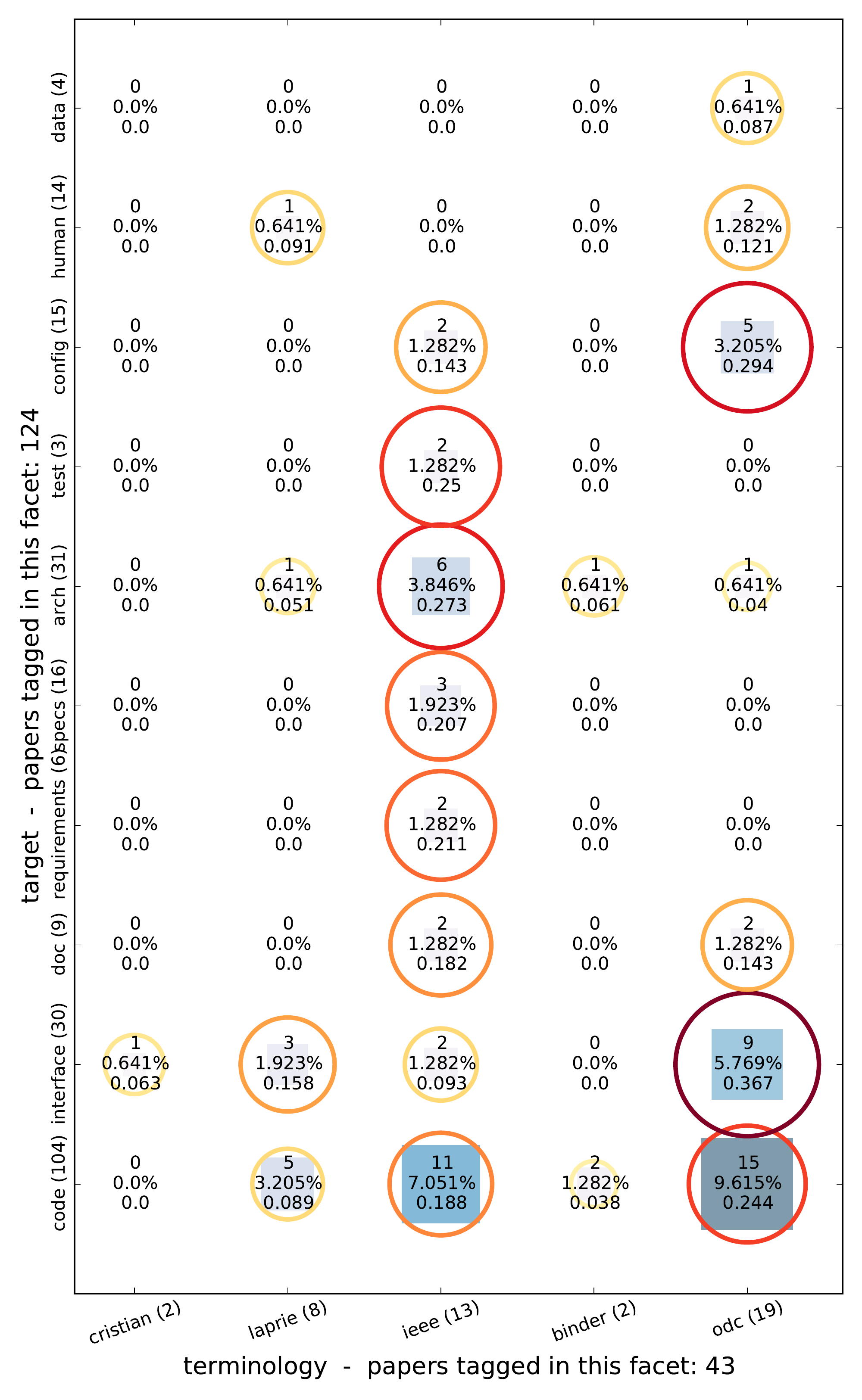}}
\caption{Occurrences of tag combinations from the terminology and target categories. The ODC terminology is mainly used to describe code and interface problems, whereas the usage of the IEEE terminology is more widespread across targets such as architecture, test and specification. \interpretation{ODC is tailored for code aspects mainly. The code and interface targets are well covered by ODC, as they are explicitly described by the ODC \enquote{Target} and \enquote{Defect Type} categories. Testing and requirements are also discussed in the ODC papers, but mainly in the context of how a bug can be reproduced or by mapping code defects to requirement. On the other hand, the IEEE has standardized terminology not only for software faults, but also for many other aspects in the whole software development life cycle. This might explain why the IEEE terminology is used for a broader range of targets.}}
\label{scatter_target_terminology}
\end{figure}

\begin{figure}[h!]
\centerline{\includegraphics[width=.85\textwidth,height=.85\textheight,keepaspectratio]{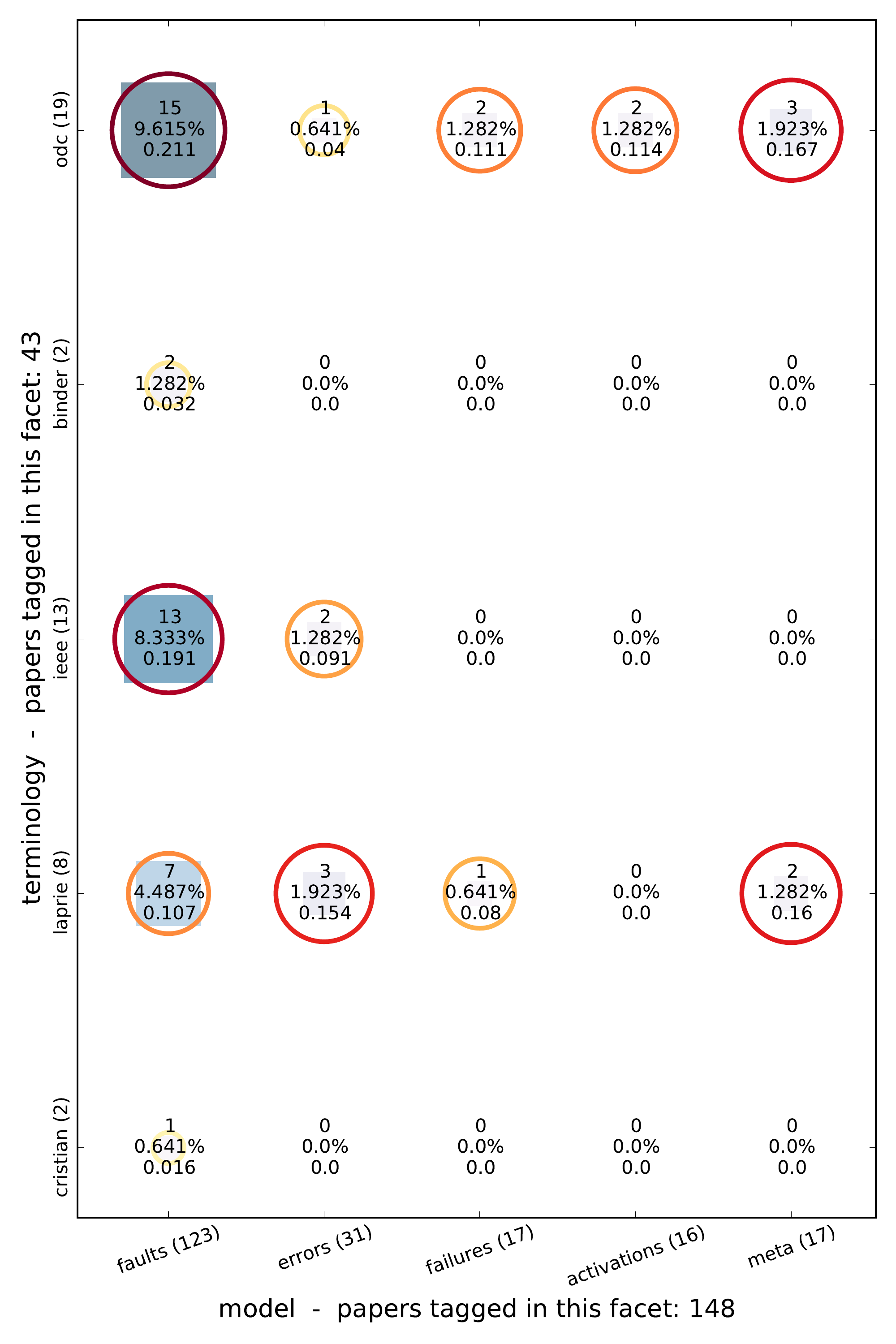}}
\caption{Occurrences of tag combinations from the model and terminology categories. The diverging sets of terminology appear to be used for different types of models: while ODC and IEEE are strong with fault models, the Laprie terminology also frequently used for error models, as well as meta studies. ODC appears to have a broader range of applicability than the IEEE terminology. \interpretation{The definitions of error states and failure events in ODC are based on Laprie, and therefore use a precise state-based definition. On the other hand, IEEE defined \enquote{fault} (error in our terminology) as a \enquote{manifestation of an error in software}, which is rather vague. It may be the case that IEEE terminology is tailored mainly to suit mainly static assets.}}
\label{scatter_model_terminology}
\end{figure}

%\begin{figure}[h!]
%\centerline{\includegraphics[width=0.9\linewidth]{../data/images/scatter_terminology_domain.pdf}}
%\caption{Occurrrences of tag combinations from the terminology and domain categories. }
%\label{scatter_terminology_domain}
%\end{figure}

\subsection{Discussion}

In this section, we present selected views on the data using our customized bubble plot visualization, in order to answer the research questions of our study (Section \ref{methodology}).

While all 2-combinations of tag categories are certainly worth studying, Figures \ref{scatter_target_model} to \ref{scatter_model_terminology} (appendix) show a selection of the plots yielding the most insights. For ease of reading, the analysis and interpretation per plot is attached to the particular graph. 

\textbf{Research Question \ref{whatclasses}} targeted the different classes of software failure causes. 
Here, it is eye-catching that a majority of the papers -- 123 out of 156 publications -- discusses a fault model rather than an error, activation or failure model.
This also implies that much research has focussed on static code features rather than dynamic phenomena such as activation patterns and error states. The prevalence of \enquote{code} in the target category (see Figure \ref{hist_all_tags}) confirms this.

The fact that models of faults, which target code, dominate most of the publications we studied, also becomes visible in subsequent bubble charts such as Figure \ref{scatter_target_model}.
As Figure \ref{scatter_model_purpose} demonstrates, the majority of publications focusses on software testing (frequently automatic test case generation) and general quality improvement.

\textbf{Research Question \ref{howcategorized}} aimed at understanding with which patterns the literature classifies software failure causes.
A large portion of the papers we studied are domain-, language- and/or paradigm-agnostic. The models discussed therein are so general as to be applicable to different technologies.

With \textbf{Research Question \ref{whatterminology}}, we asked which terminology is most relevant in the research field.
A significant portion of the papers is based on the relevant basic terminology we identified during the tagging process, and for which we created the terminology category. 43 publications explicitly name and cite another terminology model as the basis of their research (see Figure \ref{hist_all_tags}).
Among these papers, the ODC terminology model \cite{chillarege1992orthogonal} is clearly the most popular.

Figures \ref{scatter_target_terminology} and \ref{scatter_model_terminology} depict how the different sets of terminology are used. Figure \ref{scatter_target_terminology} illustrates that the Laprie, IEEE and ODC terminologies are popular for varying sets of targets.
As Figure \ref{scatter_target_terminology} shows, ODC terminology predominates when code and interface issues are described. On the other hand, IEEE terminology is used for a broader range of targets, especially for architectural discussions.
Although we have noted that ODC is used primarily to discuss code features, Figure \ref{scatter_model_terminology} shows that it can nevertheless serve for models of dynamic aspects. Failure, error, and fault activation models have also been described using ODC terminology.

We were surprised that the traditional reliability terminologies by Cristian \cite{cristian1987exception} and Laprie/\avz et al. \cite{avizienis2004basic} play only a secondary role when classifying software failures.

\textbf{Research Question \ref{whattrends}} aimed at identifying trends and research gaps.
As mentioned previously, automated software testing and fault models for this purpose seem to be trending topics in the scientific community (see Figure \ref{hist_all_tags}). 

We believe that research gaps exist in the areas of fault activation models and error models for recovery and prediction (Figures \ref{scatter_domain_purpose} and \ref{scatter_model_purpose}). As fault activation and potential error states depend largely on the runtime environment, we would have expected an amount of domain- or language-specific research in these areas. However, there is only a handful of such research papers (see Figures \ref{scatter_language_model} and \ref{scatter_model_purpose}).

\section{Related Work}
Walia et al. \cite{Walia:2009:SLR:1539052.1539605} have conducted a systematic literature review to describe and classify requirement errors. One insight from their extensive study of 149 papers provides further motivation for our research. This text uses the IEEE terminology, i.e., \enquote{fault} corresponds to \enquote{error} in our sense, and \enquote{error} describes the human action which caused the issue:
\begin{quote}
While fault classification taxonomies have proven beneficial, faults still occur. Therefore, to provide more insight into the faults, research needs to focus on understanding the sources of the faults rather than just the faults themselves. In other words, focus on the errors that caused the faults.
\end{quote}

Krsul \cite{krsul1998software} provides a thorough discussion of software security flaws -- mainly defects/faults in our sense -- and derives an own categorization from it. The work is based on a selection of previous software flaw taxonomies.

Barney et al. \cite{Barney:2012:SQT:2206429.2206498} present a systematic map comprising 179 papers on the topic of trade-offs in software quality. Here, a \gls{sms} process with multiple reviewers involved is also described. Similarly to our experiences, the authors noticed that an iterative refining of tag categories and tag names is necessary.

A comparative study of how the \gls{sms} process has been conducted with multiple reviewers, as was the case in this study, is presented in \cite{6092586}. One result of this study was that although the \gls{sms} process is meant to be formally well defined, the paper selection and tagging strategies differ across teams. This fits our impression that there are still some subjective and underspecified aspects of the \gls{sms} process.

The Common Weakness Enumeration (CWE) \cite{christey2013common} is a community effort of classifying software faults based on taxonomies from the security domain. It contains anecdotal evidence from failures of real-world software systems. Weaknesses are, amongst other categorizations, classified according to affected programming languages, their severity, and the kinds of error states they cause.

\section{Future Work}

The research field of software dependability is vast and heterogeneous depending on which purpose or domain is targeted.
We believe that more meta-studies could facilitate understanding and guide the focus of further research.
In particular, follow-up studies should be done to address some further research questions on the basis of our results:

Foremost, it should be investigated whether the studied models of software failure causes match observations from real world systems.
This clearly demands for up-to-date failure data about real software failures. The CWE database could yield suitable starting points for such studies.

Also, the question arises whether current software dependability tools fit the prominent software failure cause models.
Further literature studies can help to understand the state of the art dependability tools such as static verifiers, test case generators, fault injectors, monitoring and automated error recovery solutions.

We intend to further research how fault injection approaches can benefit from the results of our study -- especially, how future models focussing on fault activation conditions can increase fault injection precision and coverage.

\section{Conclusion}

We presented a categorization of relevant papers and a structuring of published software failure cause models.
The systematic mapping process allowed us to identify 156 relevant research articles discussing software failure causes in a repeatable fashion. This vast amount of available software failure cause models was structured using eight tag categories and evaluated to observe trending topics and research gaps.

With the results presented in this study, we hope to encourage further discussion, especially from practical viewpoints, about whether the landscape of software failure causes as we have observed it, is accurate. On the basis of our systematic map, practicioners as well as researchers can identify the relevant existing literature cluster concerning any subfield of software failure cause modelling and use it to target further efforts.

One striking insight was that the Laprie/\avz terminology which stems from the traditional reliability engineering domain, is not very popular for newer software-focussed research. Instead, the IEEE definitions and the ODC terminology concept are used widely, as they provide fine-grained vocabulary for describing software faults which can be incorporated into the development and quality assurance process.

Our most crucial observation is that a majority of the research focusses on fault models by discussing source code defects. The automated software testing community is well represented.
However, there are limits on the coverage of purely code-based software dependability approaches. Many failures depend on the timing behaviour, configuration and environment of the system. Certain bugs do not cause visible failures at once, but lead to error states which can accumulate further.
It is widely known that software failures frequently have complex causes, which has led to the coining of terms such as \enquote{Heisenbug}, \enquote{software ageing}, \enquote{Mandelbug} and \enquote{Schr\"odingbug} \cite{trivedi,gray1986computers,parnas1994software}.

Notwithstanding the common awareness that static fault models do not capture the details of many software failures, our meta study shows that few models of fault activation conditions and error states have been presented in the literature. These models, which take runtime features into account, are mainly used for error recovery.
We conclude that the research on software fault models needs to be extended to better accommodate dynamic aspects such as fault activation patterns and error states during runtime.

\clearpage
%\section*{Appendices}
\begin{sidewaystable*}[h!]
%\hbox to \linewidth{\hss\hbox{%
{\renewcommand{\arraystretch}{1.2}% for the vertical padding
\begin{tabular}{|l|p{0.6in}|>{\tiny}p{3.4in}|>{\small}p{2.4in}|}
\hline
\textbf{Database} & \textbf{Results}& \textbf{\normalsize Search String} & \textbf{Further Criteria} \\ \hline
Springer & 446 & ((("fault model" OR "error model" OR "defect model" OR "bug model" OR "fault classification" OR "error classification" OR "defect classification" OR "bug classification" OR "model of software faults" OR "model of software errors" OR "model of software defects" OR "model of software bugs" OR "fault taxonomy" OR "error taxonomy" OR "defect taxonomy" OR "bug taxonomy") AND "software" AND NOT "hardware" AND NOT "circuit" AND NOT "MATLAB" AND NOT "SIMULINK")) & constrained to content type \enquote{article} and discipline \enquote{computer science}, manually excluded many false positives before downloading citations \\

ScienceDirect & 258 & (("fault model" OR "error model" OR "defect model" OR "bug model" OR "fault classification" OR "error classification" OR "defect classification" OR "bug classification" OR "model of software faults" OR "model of software errors" OR "model of software defects" OR "model of software bugs" OR "fault taxonomy" OR "error taxonomy" OR "defect taxonomy" OR "bug taxonomy") AND "software" AND NOT "hardware" AND NOT "circuit" AND NOT "MATLAB" AND NOT "SIMULINK") & selected computer science related journals and category \enquote{computer science}\\

WebOfScience & 114 & TITLE:(((((((((((((((("fault model"  OR "error model") OR "defect model") OR  "bug model") OR "fault classification")  OR "error classification") OR "defect classification")  OR "bug classification")  OR "model of software faults") OR "model of software errors") OR "model of software defects") OR "model of software bugs") OR "fault taxonomy") OR "error taxonomy") OR "defect taxonomy") OR "bug taxonomy") AND "software") OR TOPIC:(((((((((((((((("fault model"  OR "error model") OR "defect model") OR  "bug model") OR "fault classification")  OR "error classification") OR "defect classification")  OR "bug classification")  OR "model of software faults") OR "model of software errors") OR "model of software defects") OR "model of software bugs") OR "fault taxonomy") OR "error taxonomy") OR "defect taxonomy") OR "bug taxonomy") AND "software") & constrained to content type \enquote{article} and discipline \enquote{computer science}, excluded IEEE to avoid duplicates, excluded non-related conferences and journals \\

IEEE & 326 & ((("fault model" OR "error model" OR "defect model" OR "bug model" OR "fault classification" OR "error classification" OR "defect classification" OR "bug classification" OR "model of software faults" OR "model of software errors" OR "model of software defects" OR "model of software bugs" OR "fault taxonomy" OR "error taxonomy" OR "defect taxonomy" OR "bug taxonomy") AND "software" AND NOT "hardware" AND NOT "circuit" AND NOT "MATLAB" AND NOT "SIMULINK")) &  \\

Arxiv.org & 6 & (((((((((((((((("fault model"  OR "error model") OR "defect model") OR  "bug model") OR "fault classification")  OR "error classification") OR "defect classification")  OR "bug classification")  OR "model of software faults") OR "model of software errors") OR "model of software defects") OR "model of software bugs") OR "fault taxonomy") OR "error taxonomy") OR "defect taxonomy") OR "bug taxonomy") AND "software") &  \\

ACM & 42 & (Abstract:"software") and (Abstract:"fault model" or Abstract:"error model" or Abstract:"defect model" or Abstract:"bug model" or Abstract:"fault classification" or Abstract:"error classification" or Abstract:"defect classification" or Abstract:"bug classification" or Abstract:"model of software faults" or Abstract:"model of software errors" or Abstract:"model of software defects" or Abstract:"model of software bugs" or Abstract:"fault taxonomy" or Abstract:"error taxonomy" or Abstract:"defect taxonomy" or Abstract:"bug taxonomy") & two searches: \enquote{find abstract with} (shown search strings) and \enquote{find with title} (analogously), there was no support for a NOT search field \\
\hline
\end{tabular}}
%}\hss}
\caption{Search strings for various databases and platforms, and the number of results found during primary search. Many of these publications were discarded as duplicates or irrelevant to our study later. The initial searching of the databases took place on July 22, 2014.}
\label{tab:searchstrings}
\end{sidewaystable*}

\clearpage
\section*{References}
\bibliography{mappingrelated,merged}

\end{document}